\documentclass{nature}
\usepackage{graphicx}

\title{Group-level selection avoids the tragedy of the commons} 

\author{Arend Hintze$^{1,2,4}$, Jochen Staudacher$^{3}$, Katja Gelhar$^{3}$, Alexander Pothmann$^{3}$, Juliana Rasch$^{3}$ \& Daniel Wildegger$^{3}$}

\begin{document}

\maketitle

\begin{affiliations}
 \item Dalarna University, Institute for Complex Dynamical Systems and MicroData Analytics, Sweden
 \item BEACON Center for the Study of Evolution in Action, Michigan State University, USA
 \item Kempten University of Applied Sciences, Faculty of Computer Science, Germany
 \item corresponding author
\end{affiliations}

\begin{abstract}
The public goods game is a famous example illustrating the tragedy of the commons\cite{hardin2009tragedy}. In this game cooperating individuals contribute to a pool, which in turn is distributed to all members of the group, including defectors who reap the same rewards as cooperators without having made a contribution before. The question is now, how to incentivize group members to all cooperate as it maximizes the common good. While costly punishment\cite{helbing2010punish} presents one such method, the cost of punishment still reduces the common good. Here we show how group-level selection can be such an incentive, and specifically how even fractions of group-level selection can overcome the benefits defectors receive.  Further, we show how punishment and group-level selection interact. This work suggests that a redistribution similar to a basic income that is coupled to the economic success of the entire group could overcome the tragedy of the commons.
\end{abstract}

\section{Introduction}
The \textit{tragedy of the commons}\cite{hardin2009tragedy} is a well studied model in which the interests of the group are pitched against the interests of the individuals. Individuals either contribute to the common good (cooperate), or withhold their investment (defect). The common good can experience a growth in value due to synergy, which consequently benefits everyone, also the defectors. In the end, tragically, defectors will always receive a higher reward than the cooperators, even though a higher total gain could be achieved if everyone would cooperate in the first place.

As such, this model has been extensively studied to describe social systems, in which for example taxes represent the common good, and tax evaders would be defectors of that game. Obviously, we are interested in methods which encourage everyone to cooperate, overcoming the individual benefit gained from defecting. Many different solutions have been identified which promote cooperation, such as reciprocity~\cite{nowak2006five}, 
green beard effects~\cite{hamilton1964evolution}, or costly punishment of defectors~\cite{fehr2000cooperation,hintze2015punishment}. Similarly, we know that in games 
played spatially cooperation often dominates~\cite{helbing2010punish} compared to well mixed situations. Here we investigate the effect of group-level selection, where the payoff of the individual is not only dependent on its own choices, but also of that of the group. Group-level selection is a concept derived from nature. Evolution is normally selecting the individuals of a population according how well fit they are to their environment, as they produce the most viable offspring. However, organisms often form groups to take advantage of mutual benefits that such behavior grants. Fighting off enemies by swarming, division of labor, or other forms of collaboration come to mind. If not only individuals experience the benefit, but the group as a whole enjoys reproductive success over another group, we speak of group-level selection. Often the terms kin selection, multilevel selection, and inclusive fitness are used interchangeably\cite{kramer2016kin} even though they refer to distinct concepts. Kin selection would require the members of a group to also be selected by their genetic distance, which we do not consider here. Inclusive fitness on the other hand refers to a much larger concept. In predator prey dynamics the fitness of the prey is dependent on the fitness of the predator leading to the ``fit when rare'' phenomena for example. Group-level selection in the strictest sense requires all members to be selected and being allowed to propagate offspring into the next generation. The slime mold \textit{Dictyostelium discoideum}\cite{devreotes1989dictyostelium} and its life cycle illustrates the the difference between individual and group-level selection. In its amoeba stage, cells can replicate individually, and evolution occurs on the level of the individual. When food becomes sparse, cells aggregate and first form a mobile slug which later culminates into a fruiting body. The group of cells forming the fruiting body can now experience the rewards of group-level selection when wind disperses the spoors. The individual spores in turn become amoebas again, and so forth.  

However, it is easy to imagine a mixed model between those two extremes\cite{jahns2018integration}. Instead of selecting the entire group, the payoff each individual of the group receives, can be pooled and redistributed. Each member would receive the same share, even though the total amount distributed would still depend on the actions each individual took. Selection could then happen on an individual level, but group members would have the same chance of propagating offspring into the next generation. The pooling of resources to allow for group-level selection in the public goods game has been introduced earlier\cite{boza2010beneficial} but not its fractional redistribution. The fraction of payoffs that can be pooled or remain at the individual can be dialed. We call this fraction the degree of individualism $i$. Imagine the degree of individualism to be $i=0.5$. In that case, 50\% of the payoff each individual receives would be pooled and redistributed equally, while the other 50\% of the payoff remains with the individual without being redistributed. In the extreme case of $i=0.0$ all payoff is redistributed effectively becoming a group-level selection regime (similar to that in Boza et al.~2010~\cite{boza2010beneficial}). A level of individualism of $i=1.0$ would on the other end of the spectrum represent an individual selection regime. The expectation is that different degrees of individualism cause players to change their behavior. For example, a single defector might receive a higher payoff than a cooperator, but at the same time such behavior lowers the total payoff the group received. A lower level of individualism ($i<1.0$) now couples the payoff of said defector more tightly to the success of the group. The idea is to incentivize an individual to cooperate more by coupling its payoff to that of the group, and study under which circumstances this leads to higher degrees of cooperation. We think this mechanism could be implemented in social systems as well. Instead of an unconditional basic income, one could offer a basic income linked to the economic success of the social group. We will show how different degrees of group-level selection leads to cooperation, and show what role punishment plays in this context.

\section{Synergy Controls Cooperation in the Public Goods Games}
We analyze the public goods game following Hintze 2015~\cite{hintze2015punishment}. Each individual in a group of $k+1$ players, i.e.~the focal player and her 
$k$ participants, can either cooperate by making a contribution of 
$1$ unit to a common pool or defect and withhold that contribution.
The sum of all contributions in the common pool is multiplied by a synergy factor $r$ and and then divided equally among all participants, i.e.~cooperators and defectors alike. 
In the case $1 < r < k+1$ a dilemma arises as it is a dominant strategy for 
the individual to defect whereas mutual cooperation would be most 
beneficial for all. 
Punishment~\cite{helbing2010punish}, i.e.~providing each player 
with the option to impose a punishment fine $\beta$ on other 
players who defect, has 
been studied in order to overcome this dilemma. Punishment comes with 
a cost $\gamma$ for the punisher and we can observe four 
types of behavior. Cooperators and defectors that do not punish as well as 
moralists, i.e.~cooperators that punish defectors, and immoralists, i.e.~players who punish other defectors while simultaneously defecting themselves. 
Following~\cite{hintze2015punishment} 
$N_{C}, N_{D}, N_{M}, N_{I}$ denote the numbers of cooperators, defectors, moralists and immoralists and 
$P_{C}, P_{D}, P_{M}, P_{I}$ the corresponding payoffs.

In our approach we additionally introduce the parameter $i \in [0,1]$ as a level of individualism such that for $i=0$ only the surplus is distributed among the players and not the initial contributions of $1$ unit of cooperating players to the public good whereas for $i=1$ we end up with the original public goods game.

\subsection{Critical Points without punishment}
We first analyze a public goods game with our parameter $i$, but without punishment, i.e. for the punishment fine $\beta$ and the punishment cost $\gamma$ there holds $\beta=\gamma=0$ and hence $N_{M}=N_{I}=0$.

In this setting the payoff of a collaborator is 
\begin{equation} \label{CooperatorWithoutP}
P_{C} = i (r \frac{N_{C}+1}{k+1} -1) + (1-i) (r-1) \frac{N_{C}+1}{k+1}
\end{equation}
whereas the payoff of a defector is given by 
\begin{equation} \label{DefectorWithoutP}
P_{D} = i r \frac{N_{C}}{k+1} + (1-i) (r-1) \frac{N_{C}}{k+1}
\end{equation}

In both equations \ref{CooperatorWithoutP} and \ref{DefectorWithoutP} the first addend signifies the gains on individual level whereas the second addend stands for distribution of the gains on group-level. Note that the factor $r-1$ in the second addend reflects the fact that for cooperators only the surplus beyond the original contribution of $1$ is distributed among the group.

Equations \ref{CooperatorWithoutP} and \ref{DefectorWithoutP} can be simplified to 

\begin{equation} \label{CooperatorWithoutP1}
P_{C} = \frac{N_{C}+1}{k+1} (r+i-1) - i
\end{equation}
and 
\begin{equation} \label{DefectorWithoutP1}
P_{D} = \frac{N_{C}}{k+1} (r+i-1)
\end{equation}

In order to find the critical point $r_{C}$ we investigate $P_{C} - P_{D} > 0$ and calculate it to be 
\begin{equation} \label{criticalPointWithoutP}
r_{C} = i (k+1) - i + 1 = i k + 1.  
\end{equation}
Note that for $k=4$ expression \ref{criticalPointWithoutP} is very plausible as it means $r_{C} > 1$ for $i=0$ (i.e.~only the group-level counts) and $r_{C} > k+1 =5$ for $i=1$ (i.e.~original public goods game in which only the individual level counts) whereas we obtain $r_{C} > 3$ for $i=0.5$ (and one can easily check that in this case cooperating is still a dominant strategy even if all $k=4$ neighbours defect).

\subsection{Critical Points with punishment}
We now extend our analyses to the case of punishment, i.e. there is a punishment fine $\beta$.

Then the payoff of a collaborator is 
\begin{eqnarray*} 
P_{C} &=& i (r \frac{N_{C}+N_{M}+1}{k+1} -1) + (1-i) (r-1)   \frac{N_{C}+N_{M}+1}{k+1} - (1-i) \beta \frac{N_{M}+N_{I}}{k+1} = \\
& = & \frac{N_{C}+N_{M}+1}{k+1} (r+i-1) - i  - (1-i) \beta  \frac{N_{M}+N_{I}}{k+1}
\end{eqnarray*}
whereas the payoff of a defector is given by 
\begin{eqnarray*} 
P_{D} & = & i (r \frac{N_{C}+N_{M}}{k+1} - \beta \frac{N_{M}+N_{I}}{k+1}) + (1-i) (r-1) \frac{N_{C}+N_{M}}{k+1} - (1-i) \beta \frac{N_{M}+N_{I}}{k+1} = \\
& = & \frac{N_{C}+N_{M}}{k+1}  (r+i-1) - \beta \frac{N_{M}+N_{I}}{k+1}
\end{eqnarray*}

Following \cite{hintze2015punishment} we abbreviate $\rho_{P} = \frac{N_{M}+N_{I}}{k+1}$ and interpret it as the density of punishers.

In order to find the critical point $r_{C}$ we investigate $P_{C} - P_{D} > 0$ and calculate it to be 
\begin{equation} \label{criticalPointWithP}
r_{C} = i (1 - \beta \rho_{P}) (k+1) - i + 1 
\end{equation}
Note that this results is very plausible as it means $r_{C} > 1$ for $i=0$ (i.e.~only the group-level counts) and $r_{C} > (k+1) \cdot (1 - \beta \rho_{P})$ for $i=1$ (i.e.~reproducing the findings from equation (14) from the paper \cite{hintze2015punishment}) whereas we obtain our original expression \ref{criticalPointWithoutP} for $\beta =0$.

We finally observe that for $i \in [0,1[$ our approach incorporates an implicit punishment cost for the cooperators even in the case $\gamma = 0$ as punishment reduces the group-level payoff for both the cooperators and the defectors. Note also that in any case any punished defectors pay their punishment fine $\beta$ completely for any value $i \in [0,1]$.

\section{Computational Evolutionary Model}
The above described mathematical formalism should translate to a system of evolving agents. However, our mathematical formalism assumes infinite populations and their individual actions to be equivalent to mean play frequencies. Also, evolution is defined as a process comprised of inheritance, variation, and selection. In order to replicate this process accurately, we can not disregard the effect of stochastic and discrete mutations, as we would otherwise not model evolutionary but only population dynamics~\cite{adami2016evolutionary}. Therefore, we confirm our above findings by using a computational agent based evolutionary model. A population of agents plays the above described public goods games and their performances define their reproductive success. The actions of each agents are encoded by genes, specifically by a pair of probabilities. These probabilities define each agent's likelihood to cooperate $p_{C}$ and to punish $p_{P}$. When agents are selected to transmit offspring into the next generation, their genes (probabilities) can experience mutations. This model is identical to the one used in Hintze 2015~\cite{hintze2015punishment} except for the payoff function. Instead, payoffs are now distributed among the members of the group depending on the degree of individualism $i$. In the case of $i=1$ each player's payoff is independent of the group payoff and thus identical with the classic public goods game. In the case of $i=0$ the reward of the individual is the average payoff of the group (as defined before). We also know that the public goods game is dependent on the synergy factor $r$ as well as the punishment fine $\beta$ and the punishment cost $\gamma$. This technically creates a four dimensional parameter space with the axes $i$ (in $[0,1]$), $r$, $\beta$, and $\gamma$ (all three in $[0,\infty[$).

\subsection{Without Punishment}
\begin{figure}[htb]
    \centering
    \includegraphics[width=0.9\textwidth]{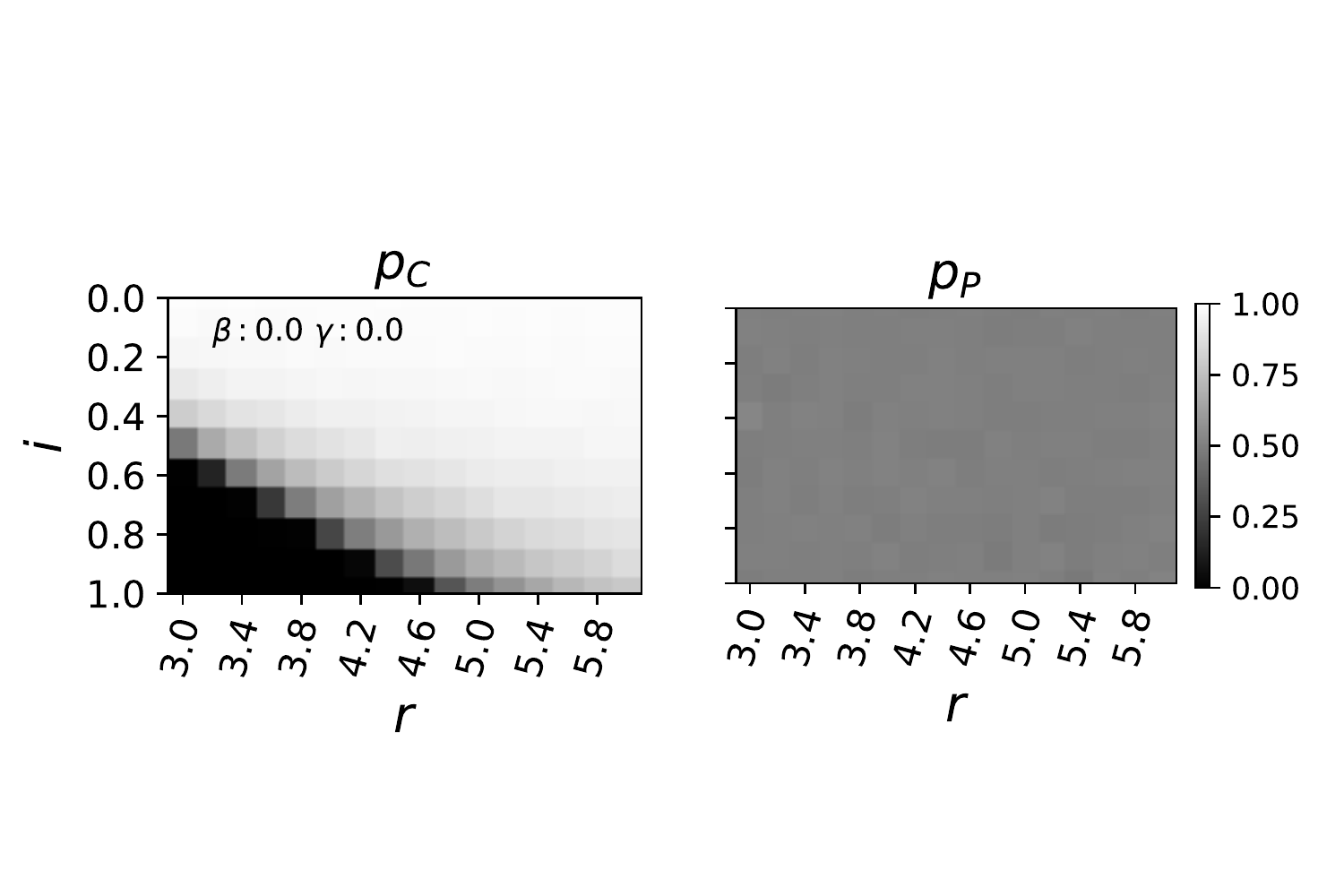}
    \caption{Phase diagram for evolved strategies under varying conditions. In order to model the absence of punishment $\beta$ and $\gamma$ were set to $0.0$ while the synergy factor $r$ (x-axis) and the degree to which payoffs were distributed individually $i$ (y-axis) were varied. On the left the probability to cooperate at or after the point of convergence is shown, on the right the probability to punish. The color bar on the right shows that probabilities of $0.0$ are black, and probabilities of $1.0$ are displayed in white. For each square in the phase diagram $100$ replicate evolutionary runs over $100.000$ generation were performed.}
    \label{fig:zeroCost}
\end{figure}

To model a game played without punishment we set $\beta=0$ and $\gamma=0$, and explore only the parameter space for the synergy factor $r$ from $3$ to $6$ (in increments of $0.2$) as well as the level of individual payoff $i$ from $0$ to $1$ (in increments of $0.1$). For each combination of factors, we ran $100$ independent replicate experiments for $100.000$ generations. After the line of descent~\cite{lenski2003evolutionary} was reconstructed, the final $100$ generations from all replicate runs were averaged to determine the point of convergence. We find the predictions about the critical points without punishment confirmed (see Figure \ref{fig:zeroCost}). For $i=1.0$ (no group-level selection) we find the critical point for strategies to cooperate at $r=5$, i.e.~$k+1=5$. Below that we find strict defection and cooperation to start above that critical point. With an increase of group-level selection ($i<1$) we find the critical point to move to a lower $r$. As such, cooperation has it easier to evolve the more group-level selection happens. The gene for punishment, as it is neither costly nor rewarding ($\beta=0$ and $\gamma=0$), is drifting, indicated by the genes average value to converge on $0.5$.

\subsection{With Punishment}
To confirm the effects of punishment and group-level selection in the public goods game, six different combinations of cost and fine were tested. Again $100$ replicate evolutionary experiments per parameter combination were run and analyzed as before. 

\begin{figure}[htb]
    \centering
    \includegraphics[width=0.4\textwidth]{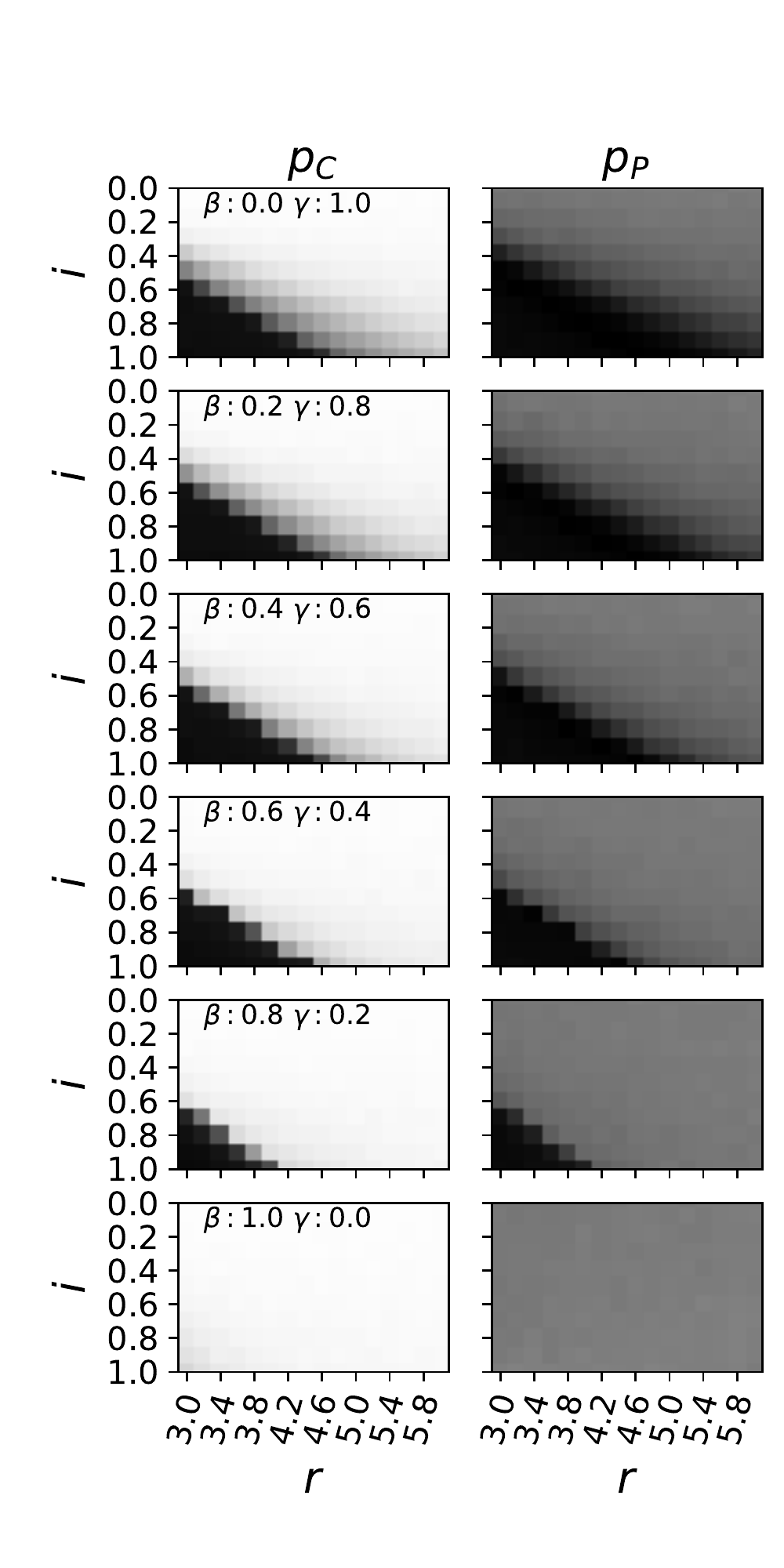}
    \caption{Phase diagrams for evolved strategies under varying conditions for punishment. From top to bottom: The punishment fine $\beta$ was increased over six experiments from $0.0$ to $1.0$, while at the same time the cost of punishment $\gamma$ was reduced from $1.0$ to $0.0$. The left column shows the probabilities to cooperate after evolution converged, on the right the same for the probability to punish. Everything else is identical to figure \ref{fig:zeroCost}.}
    \label{fig:bigSweep}
\end{figure}

As predicted by the mathematical model, we find the critical point at which cooperation starts to emerge to be dependent on the degree of group-level selection $i$ as well as on punishment, i.e.~the punishment fine $\beta$ 
and the punishment cost $\gamma$ (see Figure \ref{fig:bigSweep}). With an increase of the punishment efficiency, controlled by an increase of the fine $\beta$ and a decrease of the cost for that fine $\gamma$, we find less synergy $r$ to be necessary for cooperation to evolve. Similarly, we also observe that evolved strategies do not punish when they also do not cooperate. Consequently, when they do cooperate, the punishment gene starts to drift ($p_{p}=0.5$) as observed before~\cite{hintze2015punishment}. When all strategies become cooperators and no one punishes, punishment does not happen, and thus no cost is applied, explaining why the punishment gene drifts under those conditions.

In the case where punishment is not costly anymore ($\gamma=0.0$) we find all strategies to become cooperators, and the punishment gene is under drift again.

\section{Discussion}
We introduced a new way to redistribute the payoff in a group of players participating in the public goods game. In the case of individual level selection, if the synergy between the players is low, defection becomes the optimal strategy. In the case of pure group-level selection on the other hand, the total payoff the group receives dictates cooperative behavior. The important question answered here is whether resources can be distributed differently and in such a way that individual actions still affect the payoff of the individual while simultaneously coupling the payoff of the individual to the accomplishments of the group. The redistribution of resources according to the degree of individualism $i$ allows for this to happen. We showed mathematically and by using a computational evolutionary model that this form of redistribution indeed promotes cooperation. The lower the degree of individualism, the sooner group members start to cooperate. 

Costly punishment has been identified as an alternative factor that also promotes cooperation. We found that to be true, and also that costly punishment has a synergistic effect when combined with higher levels of group-level selection. However, we also found that the degree of individualism seems to be a much better way to promote cooperation. When $i<0.5$ we find cooperation ubiquitously present regardless of costly punishment, something only achieved without group-level selection when punishment becomes free ($\beta=1.0$ and $\gamma=0.0$).

A similar argument is believed to be an important driver for the economy: people are most motivated when their efforts translate into individual gains. Here we showed, that full cooperation, and thus the remedy to the \textit{tragedy of the commons} can already be achieved at $i<0.5$. At higher levels of individualism, costly punishment can be used to achieve the same. This suggests that high levels of taxes combined with an equally fair redistribution of wealth, for example due to a basic income, foster cooperation without the need for costly punishment. Or if a higher degree of individualism is desired, punishment can be used as well to promote cooperation, and thus higher total payoffs.

Interestingly, as soon as $i<1$ any punishment fine $\beta$ to be paid 
in full by punished defectors leads to an implicit punishment cost for any cooperating individuals, too. In our theoretical analysis we recognized 
this implicit punishment cost 
to be a constant fraction, i.e.~$1-i$ times the punishment fines paid by the 
punished defectors. One may interpret there is less to be distributed 
within a society that 
opts for e.g.~a larger police force or longer prison sentences. In terms of 
modeling public goods games our findings imply that the punishment fine 
$\beta$ is the more important parameter as compared to the punishment cost 
$\gamma$ which may safely be forgone unless these costs are desired to be disproportional to the fine $\beta$.

In conclusion, a better redistribution of resources can make the tragedy of the commons obsolete. While there are many ways to facilitate this, a basic income that is coupled to the gross domestic product could be one way.

\bibliographystyle{naturemag}
\bibliography{sample}


\begin{addendum}
 \item This research was partially funded by the Bavarian State Ministry of Science and Arts. The authors wish to thank Fabian Glei\ss ner, currently a student at Kempten University of Applied Sciences, for his contributions during the very early stages of this research project.
 \item[Competing Interests] The authors declare that they have no
competing financial interests.
 \item[Correspondence] Correspondence and requests for materials
should be addressed to Arend Hintze~(email: ahz@du.se).
\end{addendum}


\end{document}